\documentclass[12pt,a4paper,]{article}
\usepackage{amsmath,amssymb,amsfonts,amsthm}

\usepackage[latin1]{inputenc}
\usepackage[english]{babel}
\usepackage{graphics}
\usepackage{amsfonts}

\begin{document}

\title{Noncommutative Geometry, Topology \\ and the Standard Model Vacuum}

\author{R. A. Dawe Martins \\ Nottingham University, University Park, \\ Nottingham, NG7 2RD,
UK \\ rachel.dawe@maths.nottingham.ac.uk}

      \maketitle

\begin{abstract} As a ramification of a motivational discussion for previous joint work, in
which equations of motion for the finite spectral action of the Standard Model were derived, we
provide a new analysis of the results of the calculations therein, switching from the perspective
of Spectral triple to that of Fredholm module and thus from the analogy with Riemannian geometry to
the pre-metrical structure of the Noncommutative geometry. Using a suggested Noncommutative version
of Morse theory together with algebraic $K$-theory to analyse the vacuum solutions, the first two
summands of the algebra for the finite triple of the Standard Model arise up to Morita equivalence.
We also demonstrate a new vacuum solution whose features are compatible with the physical mass
matrix. \end{abstract}

\pagebreak[4]

\section*{Introduction}

This article is a continuation  of previous work joint with Barrett \cite{smv} in which
field equations were calculated for the full set of internal space metric fluctuations allowed by
the Noncommutative geometry axioms in the spectral triple formulation of the standard model. We
begin with a development of the discussion begun in the previous work, and then provide a new
analysis of the results of the calculations therein from the perspective of Fredholm module instead
of Spectral triple. Studying these Fredholm modules using algebraic $K$-theory and $K$-homology
leads to a suggested Noncommutative version of Morse theory, - a well-known tool for studying the
topology of manifolds - which is applied to the finite spectral action.

As this work ramifies from \cite{smv}, for this article to make sense it is necessary to give an
explanation of the key concepts of the previous work before the main analysis in this article can
begin. Furthermore, discussion given in the previous work is brief and so one of the purposes
of this paper is to explain how it highlights an open question about the Noncommutative framework.
This explanation leads into a detailed introduction to the main analysis given at the beginning
of the main section, entitled `Fredholm module solutions'.

    \subsection*{Context}

More details about the tools and formalisms referred to below are provided in the background.

The context of \cite{smv} and this  its `second chapter', is on the spectral action principle
\cite{sap} by Connes and Chamseddine, where the standard model is formulated with a product (whose
image is called the total space) of two spectral triples:- one that represents the Euclidean
space-time manifold and the other the 0-dimensional internal space of particle charges. The
space-time coordinate functions remain commutative but the internal space is a noncommutative
`manifold'. The spectral action principle is an important step towards the unification of gravity
with particle physics;  the Einstein-Hilbert action plus Weinberg-Glashow-Salam theory all result
from a calculation of the eigenvalues of the Dirac operator on the total space and since the
Dirac operator encodes the metric, the spectral action principle is a purely geometrical
theory.

The scope of this study involves irreducible finite real spectral triples over the complex numbers;
irreducible in the sense that there is no proper invariant subspace of the Hilbert space for which
the triple restricted to that space is itself a (non-degenerate) triple \cite{Schucker}. For
example, the standard model finite triple of three fermion families is reducible whereas the one
family triple is irreducible. By finite, we mean a finite dimensional Hilbert space over a
semi-simple algebra. A caveat is that calculations carried out in this and the previous article
apply only in the context of these 0-dimensional geometries with Euclidean signature. This means
that at present, no direct physical inference can be made.

Currently concerned with the internal space triple by itself and not the full standard model
tensor product triple, we consider a single point of space-time: we remove all terms that do not
depend solely on the fluctuations of the internal space Dirac operator $D_F$. The finite spectral
action corresponds to the Higgs potential: $\mathrm{Tr}( D_F^4 -2 D_F^2)$.

The extra Einstein's equations for internal space were calculated by Sch\"ucker et al
\cite{Schucker} (for one generation of elementary fermions) by minimising the Higgs potential with
respect to the `fluctuated Dirac operator' \cite{gravity and matter}. They found that the standard
model Dirac operator was a solution. The construction of the fluctuated Dirac operator is carried
out by beginning with a choice of initial Dirac operator (to correspond to the standard model
fermion mass matrix) to satisfy the Noncommutative geometry axioms, and in analogy with the
equivalence principle, fluctuating it with the standard model's internal space algebra of
coordinates. In this way, the Higgs force is treated as an internal space version of gravity.

\section*{Background}

    \subsubsection*{Standard model finite spectral triple}

A spectral triple $(\mathcal{A},\mathcal{H},\mathcal{D})$ provides the analogue of a Riemannian
\footnote{Note: Riemannian not pseudo-Riemannian; applications are to Euclidean not Lorentzian
space-times.} spin manifold to Noncommutative geometry  \cite{gravity and matter}, \cite{Connes'
book}. It consists of a real, involutive, not necessarily commutative algebra $\mathcal{A}$, a
Hilbert space $\mathcal{H}$: a finitely generated projective module, on which the algebra is
represented, and a Dirac operator $\mathcal{D}$ that gives a notion of distance, and from which is
built a differential algebra.

 The geometry of any closed (even dimensional) Riemannian spin manifold can be fully described by a
(real and even) spectral triple (according to the reconstruction theorem) and a
noncommutative geometry is essentially the same structure but with the generalisation that the
algebra of coordinates are allowed to be non-commuting \cite{Landi}, \cite{not very basic}.

For the standard model the internal Hilbert space is: $\mathcal{H}
= \mathcal{H}_L \oplus \mathcal{H}_R \oplus \mathcal{H}_L^c \oplus \mathcal{H}_R^c$, where

\begin{displaymath}
   \mathcal{H}_L = ( \mathbb{C}^2 \otimes \mathbb{C}^N \otimes \mathbb{C}^3 )
   \oplus ( \mathbb{C}^2 \otimes \mathbb{C}^N )
\end{displaymath}

\begin{displaymath}
 \mathcal{H}_R = ( (\mathbb{C} \oplus \mathbb{C} ) \otimes \mathbb{C}^N \otimes \mathbb{C}^3 )
   \oplus ( \mathbb{C} \otimes \mathbb{C}^N )
\end{displaymath}

and whose basis is labelled by the elementary fermions and their antiparticles \cite{forces}. The
symbol $c$ is used to indicate the section represented by the antiparticles. The even triple has
the $\mathbb{Z}/2$-grading operator $\chi$, the chirality (eigenvalues +1 or -1). In either case
of $\mathcal{H}_L$ and $\mathcal{H}_R$, the first direct summand is the quarks and the second, the
leptons. $N$ stands for the number of generations.  For example, the left-handed up and down quarks
form an isospin doublet and their right-handed counterparts are singlets and there are three colours
for quarks and none for leptons. The charges on the particles are identified by the faithful
representation of the algebra on the Hilbert space.

In the definition of $\mathcal{H}$ above we see a second $\mathbb{Z}/2$-grading that splits the
Hilbert space into two orthogonal subspaces for particles and antiparticles: $\mathcal{H}^+ \oplus
\mathcal{H}^-$ or $\mathcal{H} \oplus \mathcal{H}^c$ \cite{non-com and reality}. This is called
$S^o$-reality  and is not an axiom but applies to the standard model as it excludes Majorana
masses. The $S^o$-reality grading operator $\epsilon$ satisfies: $[\mathcal{D}, \epsilon]=0$,
$[J,\epsilon]_+=0$, $\epsilon^{\ast} = \epsilon$, $\epsilon^2=1$. (Compare with reality operator $J$
explanation below.)

Let $D_F$ denote the Dirac operator that acts on the finite dimensional  internal Hilbert
space; it is the internal space counterpart of the Dirac operator that acts on
space-time. $D_F$ is a matrix whose parameters are given by the Higgs field,
Cabbibo-Kobayashi-Maskawa family mixing matrix and the Yukawa couplings \cite{sap}. In other words,
it provides the fermion mass matrix.

The \emph{choice} made for $D_F$ in order that the spectral action principle
reproduces the standard model is:

\begin{equation}        \label{ad hoc choice}
D_F =
\left(   \begin{array}{cccc}
0         &         M^{\ast}    &   0        &   0\\
M         &          0          &   0        &   0\\
0         &          0          &   0        & M^T\\
0         &          0          &   \bar{M}  &  0
\end{array}  \right)
\end{equation}

with basis left, right, then antiparticles left and right. $M = M_Q \otimes 1_3 \oplus M_L$ , and

\begin{displaymath}
M_Q =
\left(   \begin{array}{cc}
k_u \phi_1              &        k_d \phi_2\\
-k_u \bar{\phi_2}      &        k_d \bar{\phi_1}
\end{array}  \right)
\end{displaymath}

\begin{displaymath}
M_L = \left( \begin{array}{cc}
k_e \phi_1          &     k_e \phi_2  \\
0                   &     0
\end{array} \right)
\end{displaymath}

(An extra row is added to $M_L$ here so that the matrices are square, this is not normally
done and relative to other literature, the labelling $M$ is swapped with $M^{\ast}$.)

with
\begin{displaymath}
k_u =
\left(   \begin{array}{ccc}
m_u   &     0     &    0   \\
0     &    m_c    &    0   \\
0     &     0     &     m_t
\end{array}  \right)
\qquad k_d =
V_{CKM}
\left(   \begin{array}{ccc}
m_d   &     0     &    0   \\
0     &    m_s    &    0   \\
0     &     0     &     m_b
\end{array}  \right)
\end{displaymath}

\begin{displaymath}
k_e =
\left(   \begin{array}{ccc}
m_e   &     0     &    0   \\
0     &    m_{\mu}    &    0   \\
0     &     0     &     m_{\tau}
\end{array}  \right)
\end{displaymath}

$T$ denotes transposition, $\ast$ denotes hermitian conjugation, bar denotes complex conjugation,
$m_x$ are the Yukawa couplings of the elementary fermions, $V_{CKM}$ is the
Cabibbo-Kobayashi-Maskawa generation mixing matrix. $(\phi_1, \phi_2)^T$ is the (Higgs) scalar
doublet.

The finite spectral action corresponds to the Higgs potential: $\mathrm{Tr}( D_F^4 -2 D_F^2)$. If
the $\mathrm{Tr}(I)$ term is included \cite{sap}, which obviously does not affect the equations of
motion, then the action can be written $\mathrm{Tr}(MM^{\ast}-I)^2$.

 The spectral triple algebra $\mathcal{A}$ is a subalgebra of the bounded  operators on the Hilbert
space, it is a $\ast$-algebra not necessarily a $C^{\ast}$-algebra but its norm-closure in the
Hilbert space is a $C^{\ast}$-algebra. The standard model tensor product algebra is `almost
commutative':

\begin{equation}
   \mathcal{A} = C^{\infty}(M) \otimes ( \mathbb{H}  \oplus  \mathbb{C}  \oplus  M_3(\mathbb{C} ) )
\end{equation}

where the first factor is the (commutative) algebra of function on (Euclidean)  space-time and the
second factor is internal space algebra of particle charges.

The (faithful) representation $\rho$ of the finite space algebra has been worked out by Connes to
correspond to the particle charges see \cite{sap}. The first and second summand acts on the
particles while the third summand acts on the antiparticles. The basis is given by the Hilbert
space above.

\begin{equation}    \label{rho}
\rho:=
\left( \begin{array}{cccc}
\rho_L   &       0     &       0       &     0  \\
0        &    \rho_R   &       0       &     0  \\
0        &      0      &    \rho^c   &     0  \\
0        &      0      &       0       &  \rho^c
\end{array}   \right)
\end{equation}

    \subsubsection*{Real structure and Poincar\'e duality}

Instead of splitting $\mathbb{C}$ into two copies of $\mathbb{R}$ as its name  suggests, $J$ forms
two subspaces of the Hilbert space (not orthogonal as in $S^o$-reality), which are  interpreted
as fermions and antifermions, in which case $J$ is given by the composition of the charge
conjugation operator with complex conjugation. The mathematical purpose of the $J$ operator
entering the axioms is to provide Connes' notion of a `noncommutative manifold'. That is, by
turning the Hilbert space into a bimodule, the `real' structure (\cite{non-com and reality}) allows
the generalisation of Poincar\'e duality to the spectral triple; $a \in \mathcal{A}$ and the
opposite algebra, $b^o \in \mathcal{A}^o$ (or $\mathcal{A}^{op}$) where $a$ acts on the left of
$\mathcal{H}$ and $b^o$ acts on the right.

\begin{displaymath}
[a,b^o]=0, \qquad b^o = J b^{\ast} J^{-1} \quad \forall a, b \in \mathcal{A}
\end{displaymath}

The opposite algebra provides the `dual' to the algebra. The pairing of the $K$-theories of these
two algebras provides a noncommutative geometric version of Poincar\'e duality. This structure
is also important in the notion of first order differential operator in noncommutative geometry.
The tangent space over any manifold is real, and the reality structure gives rise to the
real-$K$-theory of the enveloping algebra $\mathcal{A} \otimes \mathcal{A}^o$.

The action of $J$ on $\mathcal{H}$ as given by the composition of charge conjugation and complex
conjugation:

\begin{displaymath}
   J \binom{ \psi_1}{\bar{\psi_2}}=\binom{\psi_2}{\bar{\psi_1}}
        \quad (\psi_1,\bar{\psi_2}) \in  \mathcal{H} \oplus \mathcal{H}^c
\end{displaymath}

where the bar indicates complex conjugation.

\subsubsection*{$K$-theory}

$K$-theory is a generalised cohomology theory. Topological $K$-theory is the  topological invariant
that classifies the vector bundles over a given field, over a compact topological space $X$ up to
stable equivalence. It is an abelian group $K^0(X)$ generated by the isomorphism classes of vector
bundles over a given field. Addition is given by $[E]+[F]=[E \oplus F]$ where $[E]$ and $[F]$ are
isomorphism classes of vector bundles $E$ and $F$. Every element of the group is a difference:
$[E]-[F]$. The Serre-Swan theorem provides the identification of topological with algebraic
$K$-theory. That is, $K^0(X)$ is isomorphic to the algebraic $K$-theory group $K_0(C^0(X))$. The
group $K_0(A)$ for a $C^{\ast}$-algebra $A$ is generated by the projections (self-adjoint
idempotents) in $A$. These projections form an abelian semigroup rather than a group,
but the Grothendieck construction turns them into an abelian group using an equivalence relation,
which is very much analogous to the process of constructing the integers from the natural numbers
\cite{Landi's book}, \cite{W-O}.

Some rules for $K$-theory include:

$K_0(M_n(A)) = K_0(A)$ (Morita equivalent algebras have the same $K$-theory), \\
$K_0(A \oplus B) = K_0(A) \oplus K_0(B)$  where $A$ and $B$ are $C^{\ast}$-algebras.

    \subsubsection*{Fredholm modules and K-homology}

The Fredholm module is the `pre-metric' structure that is used to define the noncommutative
calculus \cite{Connes' book}. A spectral triple can be thought of as an unbounded (unless the
Hilbert space is finite dimensional) Fredholm module with Dirac operator providing a notion of
distance.

\textbf{Definition: Fredholm operator.} A Fredholm operator is a bounded operator on a Hilbert
space whose kernel and cokernel are finite dimensional and is invertible modulo compact operators.

\textbf{Definition: Fredholm module.} \cite{Connes' book}

Let $A$ be an involutive algebra (over $\mathbb{C}$). Then a Fredholm module $(H,F)$ over $A$ is
given by: 1) an involutive representative $\rho$ of $A$ in a Hilbert space $H$. 2) a Fredholm
operator $F=F^{\ast}$, $F^2=I$ on $H$ such that $[F,\rho(a)]$ is a compact operator for any $a \in
A$. \footnote{After some trivial changes that we do not need to go into details of here, Connes
ensures that the Fredholm module makes sense in finite dimensions \cite{Connes' book}.}

If there is a $\mathbb{Z}/2$ grading $\chi$, such that $\chi=\chi^{\ast}$, $\chi^2=1$ of the Hilbert
space such that a) $[\chi,\rho]=0 \quad \forall a \in A$ and b) the anticommutator $[\chi,F]_+=0$
then the Fredholm is \emph{even}.

There is a natural assignment of a Fredholm module to a spectral triple. To be precise, an
observation given in \cite{Connes' book} is that there is a canonical assignment of a Fredholm
module to a spectral triple given by $F=\mathrm{sign}(\mathcal{D})$ (that is, $\mathcal{D}=F \vert
\mathcal{D} \vert$) outside the kernel of $\mathcal{D}$. (On the finite dimensional kernel of
$\mathcal{D}$, one takes an arbitrary isometry \cite{Sitarz}.)

Kasparov's $K$-homology is the Poincar\'e dual theory to $K$-theory:- the $K$-homology (abelian)
group of a Fredholm module is given by the homotopy classes of its Fredholm operator $F$.
Let $F$ be an elliptic operator on a compact space $X$ (all such are Fredholm), then there is
an isomorphism,  index: $[X,F] \rightarrow K_0(X)$ where $[,]$ denotes homotopy equivalence
classes. Due to  Connes' construction of Poincar\'e duality for noncommutative spaces, in which the
dual to $\mathcal{A}$ is its opposite algebra $\mathcal{A}^{op}$, one can write that $[F] \in
\rho(A^{op})$ because the abelian group $K^0(A^{op})$ is generated by the minimal rank projections
of the opposite algebra $A^{op}$. \cite{Higson Roe}.

 \subsubsection*{Index and intersection form}

We recall that every finite dimensional real involutive algebra on a finite dimensional Hilbert
space \cite{Kraj} over the complex numbers is isomorphic to the direct summand
$M_{n_1}(\mathbb{C}) \oplus...\oplus M_{n_k}(\mathbb{C})$ for some integers $n_1$ to $n_k$.
Consider the Hilbert space to be made up of separate `chunks' where each is acted upon by a
different algebra summand \cite{Sitarz}:

\begin{equation}
H_{ij} = P_i \mathcal{H} P_j, \qquad  \mathcal{H} = \bigoplus_{i,j} H_{ij}
\end{equation}

where the $P_i$ or $p_i$ are projections in $\mathcal{A}$ and the $P_j$ or $Jp_jJ^{-1}$ are
projections in $\mathcal{A}^o$. The action on $H_{ij}$ from the left is $a_i \otimes 1 \otimes 1$
and the action from the right is $1 \otimes 1 \otimes a_j^T$. Let $r_{ij}$ be the number of
particles represented by $H_{ij}$ and $\chi$. The intersection form is:

\begin{equation}
\mu_{ij} = r_{ij} \chi
\end{equation}

which has non-zero determinant.

The matrix $\mu_{ij} = r_{ij} \chi$ is the same thing as the tensor product pairing of
the $K$-theory groups of the algebra $\mathcal{A}$ with its opposite algebra:

\begin{equation}
\mu_{ij} = \mathrm{Tr} \big( \chi ( \rho(p_i) J \rho(p_j) J^{-1}) \big)
\end{equation}

and we also see that:

\begin{equation}
\mu_{ij} = r_{ij} \chi = \mathrm{dim} H_{ij}^R  -  \mathrm{dim} H_{ij}^L
\end{equation}

and then by summing over all the $H_{ij}$ one arrives at right-hand side of the Fredholm index
formula:

\begin{equation}
\mathrm{Index}(P \mathcal{D}^+ P) = \mathrm{dim} \mathcal{H}_R  -  \mathrm{dim} \mathcal{H}_L
\end{equation}

where $\mathcal{D}^+ = M^{\ast}$ in our conventions for the finite triple.

\subsubsection*{The axioms}

Axioms 1, 3 and 5 are identical with those of commutative geometry. See \cite{gravity and matter}
for a full statement and explanation of the axioms.

\begin{enumerate} \item $n>0$ $ds=\mathcal{D}^{-1}$ is an infinitesimal of order $\frac{1}{n}$ where
$n$ is the dimension of the space. \item $[[\mathcal{D},a],b^0]=0$ $\forall a,b \in \mathcal{A}$. By
axiom 7 we also have: $[[\mathcal{D},b^0],a]=0$ $b^o \in \mathcal{A}^o$ opposite algebra. \item
(Smoothness) This is the algebraic formulation of smoothness of coordinates.  \item (Orientability)
There is a Hochschild cycle $c$. For $n$ even, its representation on $\mathcal{H}$ is
$\pi(c)=a^0[\mathcal{D},a^1] \ldots [\mathcal{D},a^n]$. This defines the construction of the
analogue of the differential form that does not require a previous knowledge of the tangent bundle.
If $n$ is odd, require $\pi(c)=1$. If n is even, $\pi(c)=\chi$ satisfies: $\chi=\chi^{\ast}$,
$\chi^2=1$, $\chi D = -D \chi$  \item (Finiteness and absolute continuity) The Hilbert space is a
finite, projective $\mathcal{A}$-module possessing a hermitian structure. \item There is the
Poincar\'e duality isomorphism $K_{\ast}(\mathcal{A}) \rightarrow K^{\ast}(\mathcal{A})$ where the
intersection form is nondegenerate. \item (Reality) There is an antilinear isometry $J:\mathcal{H}
\rightarrow \mathcal{H}$ with $b^0=Jb^{\ast}J^{-1}$ and $[a,b^0]=0$. The operator $J$ must
satisfy a set of further conditions, which for 0-dimensions are the following. $J^2 = I$, $JD=DJ$,
$J \chi = \chi J$.  \end{enumerate}

\section{Gravity and internal space}

In  order to motivate the main analysis of this article we give an outline of the previous work in
\cite{smv} and discuss some of its implications. We also prove a new result.

The article \cite{smv} highlights the following issue. Einstein's equations involve all fluctuations
of the space-time metric, and so if we believe that noncommutataive spectral triples are analogous
to Riemannian spin manifolds, then we should vary the finite action with respect to the most
general internal space Dirac operator allowed by the Noncommutative geometry axioms. In other
words, since Riemannian geometry gives rise to the study of gravity, we should continue to treat
the Higgs force as an internal space version of gravity by calculating the extra Einstein's
equations for the entire set of metric fluctuations. A feature of this approach is that the element
of choice is removed; the hypothesis was that the standard model fermion mass matrix would arise as
a solution of these equations of motion, just as Newton's laws of motion are selected through an
action minimisation principle, and thus the existence of classical mechanics can be explained
mathematically. The physical mass matrix did not turn out to be a solution, in fact the additional
fluctuations over-constrained the vacuum so that the solutions were completely degenerate. However,
given the logic of this idea (of Barrett's), despite giving an unphysical result, it deserves
further attention.

In response to the result, we may consider:

(a) abandoning  the paradigm that noncommutative spectral triples be completely analogous to
Riemannian geometry, and taking the Yukawa couplings to be derived from `finely-tuned' constants,

(b) proposing  that the extra fluctuations are physical, in which case additional scalar field terms
in the action are needed (in order that the mass matrix vacuum be non-degenerate)\footnote{One
replaces the action $\mathrm{Tr}(MM^{\ast}-I)^2$ with $\mathrm{Tr}(MM^{\ast} + XX^{\ast}-I)^2$ for
some matrix $X$,}, together with an additional internal space discrete version of
gravity. Such a new interaction might arise from a background source term or a twisting of the
Dirac operator,

(c) proposing an  eighth axiom for Noncommutative geometry to act as a further geometric constraint
on the Dirac operator, which might involve a definition of curvature of internal space.

Even by leaving  out the $S^o$-reality condition to increase the number of degrees of freedom in the
Dirac operator (connecting anti-particles with particles) and imposing the first order condition as
a geometric constraint upon it, did not lead to a vacuum $M$ for the physical mass matrix because
the extra fields all had zero vacuum expectation values as shown in \cite{smv}. There is another
side of the coin revealed by modifying this calculation such that those new fields are treated as
constant numbers, which means that the Yukawa couplings determine from constants, or are constants
as in the standard model: while the fermion mass ratios are not determined of course, the vacuum
solution is a non-degenerate matrix and the product $MM^{\ast}$ is diagonal whereas $M^{\ast}M$ is
not diagonal. These features are compatible with the physical mass matrix and with consideration
(a) above. After defining some notation we give the proof.

We label the additional degrees of freedom in the standard model Dirac operator on internal space
when $S^o$-reality  condition is omitted: $g, u, x, h, v, y$. In \cite{leptoquarks}
these `leptoquarks' are variable fields and effect spontaneous breaking of colour symmetry, whereas
we are treating them as constants, so here colour symmetry remains intact. There are also two more
fields $j, l$ that arise exist when $\nu_R$ is included \cite{smv}, which was done to allow the
neutrino a mass.

We use the notation: $M=M_Q \otimes 1_3 \oplus M_L$ with

\begin{displaymath}
M_Q =
\left(   \begin{array}{cc}
a              &       b\\
c              &       d
\end{array}  \right), \qquad
M_L = \left( \begin{array}{cc}
q          &     r  \\
s          &     t
\end{array} \right)
\end{displaymath}

If $M$ is a diagonal matrix then $a, d, q, t$ are interpreted as Dirac masses for the up, down,
electron and neutrino respectively.

The field equations (\cite{smv}) for the Dirac operator with leptoquarks held constant are given in
the Appendix.  With $M_Q$ invertible \footnote{substituting $ad-bc=0$ into the equations for $M_Q$
\ref{ac} to \ref{dc} gives $g=u=x=h=v=y=0$} and with all of the additional fields held constant and
$r=s=0$ by gauge freedom the equations \cite{smv} reduce to:

\begin{eqnarray}   \label{a to d colour}
3 \vert a \vert^2 +  3 \vert b \vert^2 ~ + ~ \vert g \vert^2 + \vert u \vert^2 + \vert x
\vert^2 ~ + ~ \vert h \vert^2 + \vert v \vert^2 + \vert y \vert^2 - 3 = 0 \\
\vert d \vert^2 + \vert c \vert^2 - 1 = 0 \\
a \bar{c} + \bar{b}d = 0 \\
\vert q \vert^2 ~ + ~\vert g \vert^2 + \vert u \vert^2 + \vert x \vert^2 ~ + ~ \vert j \vert^2 - 1 =
0 \\
\vert t \vert^2 ~ + ~ \vert h \vert^2 + \vert v \vert^2 + \vert y \vert^2 ~ + ~ 3\vert l \vert^2 +
\vert j \vert^2 - 1 = 0 \\
\bar{g}h + \bar{u}v + \bar{x}y=0, \quad lj=0.
\end{eqnarray}

As claimed above, these equations give $M$ a non-degenerate set of eigenvalues with $MM^{\ast}$
diagonal and $M^{\ast}M$ not diagonal. In the case of the standard model where $\nu_R=0$, the
equations are the above minus the equation involving $t$, and with $j=l=0$. (In the previous work
$g=u=x=h=v=y$ were allowed to vary, thus there were more equations and the solution was a fully
degenerate mass matrix.)

Since $\vert a \vert^2 + \vert b \vert^2$ is identified with $m_u(\vert \phi_1 \vert^2 +
\vert \phi_2 \vert^2)$, when $g, u, x, h, v, y$ are constants means that the Yukawa
couplings are determined from numbers that are constant. The conclusion is that we have found
equations of motion for which there exists a solution that is not demonstrably incompatible  with
experiment by means of an action principle in which an element of human choice is removed. The
result ironically provides a mathematical reason for the Yukawa couplings to require
fine-tuning, however options (b) and (c) are open and there is the caveat that these results apply
only to the 0-dimensional, Euclidean case. Also, there are the extra particle-antiparticle mixing
action terms \cite{leptoquarks}.

There is also a set of equations where $j$ and $l$ are allowed to vary while the
`leptoquarks' remain constant. These are the first three equations above (\ref{a to d colour})
together with:

\begin{eqnarray*}
j=0 \\
\vert q \vert^2 + \vert g \vert^2 + \vert u \vert^2 + \vert x \vert^2 - 1 = 0 \\
\vert t \vert = \vert l \vert = \frac{1}{2} \\
 \bar{g}h + \bar{u}v + \bar{x}y=0
 \end{eqnarray*}

The conclusion is the same as above.

\section{Fredholm module solutions}

The final discussion point we would like to make with regard to \cite{smv} will motivate
the calculations that follow. Below we make the observation that the vacuum solutions to the
field equations for the full set of metric fluctuations do not pertain to the spectral
triple, but rather to the pre-metric structure of the spectral triple, namely the Fredholm module.
Rather than the standard model Dirac operator being a solution, its sign is a solution. We refer to
the  observation of Connes that any spectral triple has a Fredholm module associated to it where the
Fredholm operator $F$ of the Fredholm module is identified with the sign of the Dirac operator
(outside its kernel) of the spectral triple \cite{Connes' book}. In switching focus from spectral
triple to Fredholm module, one zooms out from the geometry to the topology. Hence, instead of
hypothesising that the equations of motion single out the correct \emph{metric}, in this section we
ask if the equations of motion can give solutions which relate to \emph{topological invariants},
that is, $K$-theory and $K$-homology.

Since the $K$-homology of a Fredholm module is isomorphic to the $K$-theory of the algebra it is
over, we should observe a relationship between a given algebra and the vacuum solution. Connes'
realisation of Poincar\'e duality in Noncommutative geometry is to define the Poincar\'e dual
to be the opposite algebra. This means that the homotopy classes of the projections in the dual
algebra are identified with the $K$-homology, which we recall is given by the homotopy classes of
the Fredholm module $[F]$. For a given algebra, one may identify a corresponding Fredholm module
solution, and below we demonstrate this for the standard model and for one other
algebra. This is only an observation, but to use this framework to obtain topological data from the
vacua, we need to make the procedure unique, so that there is a one-to-one relationship between
algebra and vacuum solution. To this end, tools from Morse theory are borrowed from commutative
geometry and a way to generalise them for this Noncommutative work is suggested.

\subsection{$S^o$-real standard model finite triple vacuum}

First we recall that the most general internal space Dirac operator given the appropriate
constraints of self-adjointness, same dynamics for particles and antiparticles, orientability,
$S^o$-reality and first order condition:-

\begin{eqnarray*}
  \mathcal{D} =  \mathcal{D}^{\ast},  \qquad [\mathcal{D},J]=0, \qquad [\mathcal{D}, \chi]_+ = 0, \\
  \qquad [\mathcal{D}, \epsilon] = 0,  \qquad [[\mathcal{D},a],b^o]=0, \qquad
[[\mathcal{D},b^o],a]=0
\end{eqnarray*}

(where $[~ ,~ ]_+$ denotes the anticommutator)

is:

\begin{equation}        \label{ad hoc}
D_F =
\left(   \begin{array}{cccc}
0         &         M^{\ast}    &   0        &   0\\
M         &          0          &   0        &   0\\
0         &          0          &   0        & M^T\\
0         &          0          &   \bar{M}  &  0
\end{array}  \right)
\end{equation}

where $M = M_Q \otimes 1_3 \oplus M_L$, and we have allowed for the inclusion of $\nu_R$. To exclude
$\nu_R$ as in the standard model, we simply delete the final column from $M^{\ast}$ (or row from
$M$).

From the first order condition,
\begin{equation}
(M \rho^c - \rho^c M) \rho_L^{T'} - \rho_R^{T'}(M \rho^c - \rho^c M)=0
\end{equation}

when $\rho_L^{T'}=0$ and $\rho_R^{T'}=1$ we find that $[M, \rho^c]=0$ (the Higgs has no colour
charge) and with the standard model representation this splits $M$ up into the direct sum of quark
and lepton masses. Further than this, the mass matrix $M$ is not constrained by the first order
condition. This means that the action does not know how the algebra is represented, and hence it is
missing some geometrical information about the manifold. When $S^o$-reality is omitted some of this
information becomes available to the action and hence (when the extra fields are kept constant)
the non-degenerate mass matrix solution of the previous section arises.

As in the previous work, we drop the $S^o$-reality condition, and allow degrees of freedom in $D_F$
to vary but we leave the first order condition until after the equations of motion have been
derived; since we are to aim to develop a method to identify the algebra given a vacuum solution,
we had better omit any axioms that involve the algebra. However, we retain the condition that  $[M,
\rho^c]=0$.

The result is:

\begin{equation}    \label{D_F}
D_F =
\left(   \begin{array}{cc}
Y   &    Z   \\
\bar{Z}   &  \bar{Y}
\end{array}  \right)
\quad = \quad
\left(   \begin{array}{cccc}
0         &       M^{\ast}     &     0      &    G   \\
M         &       0            &     G^T    &    0   \\
0         &   \bar{G}          &     0      &    M^T \\
G^{\ast}  &       0            &  \bar{M}   &    0
\end{array}   \right)
\end{equation}

(where $G$ having the same dimensionality as $M$, was not a general matrix in \cite{smv} but was
constrained by the first order condition.)  Alternate blocks are zero due to the
condition $[D_F, \chi]_+=0$.  Here we are not using the first order condition, so
$G$ and $M$ are both general matrices with complex coefficients and having dimensionality depending
on the number of fermions considered.

To calculate the equations of motion we vary the finite spectral action with respect the degrees of
freedom in $D_F$ as given above, first for the $S^o$-real case and then for the non-$S^o$-real
triple. The result of the former is the same as that given in \cite{smv} but we make a new
interpretation of it.

\subsubsection{$S^o$-real triple}     \label{S part}

The action is:

\begin{equation}           \label{fsa}
  S = \mathrm{Tr}[ (D_F)^4 - 2 D_F^2]
\end{equation}

or, with the $Tr(I)$ term included:

\begin{equation}         \label{M action}
   S = \mathrm{Tr} (MM^{\ast} - I)^2
\end{equation}

(where $I$ denotes the unit matrix)

Minimising the \ref{M action} with respect to $M$ gave the very definition of partial isometry:

\begin{equation} \label{pi solution}
M^{\ast} ( M M^{\ast} - I ) = 0
\end{equation}

and hermitian conjugate.

This result means that the mass matrix that minimises the action gives each fermion an identical
mass. The new interpretation we give is that the standard model finite triple's Dirac
operator is a solution only up to its sign, and hence only up to the conformal structure of the
spectral triple, where $\mathrm{sign}D_F = \frac{D_F}{\vert D_F \vert}$. Specifically,
the vacuum solution $M_{vac}^{\ast}$ is the partial isometry in the polar decomposition of $D^+$.
This operator $\mathrm{sign}D_F$ is the Fredholm operator for the Fredholm module associated to the
standard model finite spectral triple outside the kernel of the Dirac operator (\cite{Connes'
book}).

\subsubsection{Non-$S^o$-real triple}

The action is:
\begin{equation}
  S = \mathrm{Tr}[ (D_F)^4 - 2 D_F^2]
\end{equation}

with $D_F$ given by \ref{D_F}.

Simplifying the action by using cyclicity of  trace and the fact that $\mathrm{Tr}(X) = \mathrm{Tr}
(X^T)$:

\begin{equation}            \label{M and G action}
S = \mathrm{Tr}  \big(  -2( G^{\ast} G  +  M M^{\ast}  )  +  (M M^{\ast})^2  +
(G^{\ast} G)^2 \\
+ 2( M G G^{\ast} M^{\ast}  +  M M^{\ast} G^T \bar{G}  +  M G
\bar{M} \bar{G})   \big)
\end{equation}

or:

\begin{equation}
S = \mathrm{Tr} \big( -2( G G^{\ast}  +  M^{\ast} M)  +  (M^{\ast} M)^2  +  (G
G^{\ast})^2  \\
 +  2( G G^{\ast} M^{\ast} M  +  G \bar{M} M^T
G^{\ast}  +  G \bar{M} \bar{G} M) \big)
\end{equation}

We vary the first of the above (\ref{M and G action}) with respect to $M$ and the result is:

\begin{equation}   \label{M eqn}
  M^{\ast} ( M M^{\ast} + G^T \bar{G} - I ) + G \bar{M} \bar{G} + G G^{\ast} M^{\ast} = 0
\end{equation}

and the second with respect to $G$ and the result is:

\begin{equation}    \label{G eqn}
  G^{\ast} ( G G^{\ast} + M^{\ast} M  - I) + \bar{M} \bar{G}M + \bar{M} M^T G^{\ast} = 0
\end{equation}

The field equations for $G^T$, $M^T$, $G^{\ast}$, $M^{\ast}$, $\bar{G}$ and $\bar{M}$ are just the
the transpose, hermitian conjugate or complex conjugate respectively of the above equations for $M$
and $G$.

Although there are zeroes in $D_F$ due to orientability, the equation below is the same thing as
that above, due to there being no linear terms in $M$ or $G$:

\begin{displaymath}
D_F^3=D_F
\end{displaymath}

which is of course the result of differentiating \ref{fsa} with respect to $D_F$. In other words,
simply by substituting for $D_F$ with \ref{D_F} into $D_F^3=D_F$, precisely the equations of motion
obtained above together with all their conjugate counterparts, appear.

The conclusion in this non-$S^o$-real case is the same as that in the  $S^o$-real
one, namely that the solutions are partial isometries. Here is the proof:-

First we check if the equations of motion do have any Fredholm module solutions. To do this,
we must look for solutions such that $D_{F, ~vac}^2 = I$:

\begin{equation}   \label{D_F^2}
D_{F, ~vac}^2= \left(    \begin{array}{cccc}
M^{\ast} M + G G^{\ast}    & 0 &     M^{\ast} G^T +G \bar{M}              &   0\\
    0    &        M M^{\ast} + G^T \bar{G}     & 0 &           MG + G^T M^T    \\
\bar{G} M + M^T G^{\ast}       &  0  &       \bar{G}G^T + M^T \bar{M}      &  0  \\
0  &      G^{\ast} M^{\ast} + \bar{M} \bar{G}    &  0 &       G^{\ast} G + \bar{M} M^T
\end{array}    \right)
\end{equation}

The equations \ref{M eqn} and \ref{G eqn} (and conjugates) are equivalent to
the equation $D_F^3=D_F$, and therefore we can state that the eigenvalues of the vacuum
solution for $D_F$ are all in the set $\{-1, 0, 1 \}$. Then we see that $D_{F, vac}^2$ has
eigenvalues all 1 or 0, which means that assuming it is diagonalisable, $D_{F, vac}^2 = U p
U^{\ast}$ for some unitary matrix $U$ and where $p$ is a projection, that is, $p$ satisfies $p=p^2$
and $p=p^{\ast}$. Clearly, $U p U^{\ast}$ is a projection, in other words $D_{F, vac}^2$ is a
projection, and since the Dirac operator is self-adoint, we may conclude that all the vacuum
solutions are that $D_{F, vac}$ is a partial isometry: $( D_{F, vac}^{\ast}D_{F, vac} ) ( D_{F,
vac}^{\ast} D_{F, vac} ) = D_{F, vac}^{\ast} D_{F, vac}$. A simpler way to see this is to multiply
on both sides of the equation $D_{F, vac}^3 = D_{F, vac}$ by  $D_{F, vac}$ while recalling that the
Dirac operator is self-adjoint. Notice also that for the eigenvalues of $D_{F, vac}$ that are 1,
equation \ref{D_F^2} shows that the sum of the two types of masses for each particle add up to 1
and so even if $G$ can lift the degeneracy of the Dirac mass matrix, the total mass ends up being
the same. To summarise, the Fredholm module interpretation (\ref{S part}) is again valid in this,
the non-$S^o$-real case.

\subsection{Orthogonal complements}

Although the orthogonal complement relationship between $M$ and $G$ can already
be seen from the result of the last part, below we demonstrate it explicitly and analyse the
equations \ref{G eqn} and \ref{M eqn}. This datum will be used in the next part.

Since a partial isometry is just a projection multiplied by a unitary matrix, we let $M$ be a
diagonal projection of dimensionality $n$.  (For a partial isometry $v$, $v \mathcal{H} = v
v^{\ast} \mathcal{H}$). We do not assume that $G$ is diagonal. In any calculation we assume that
the dimensionalities of the matrices $M$ and $G$ are the same.

From \ref{M eqn} and \ref{G eqn} it is immediately clear that when $M=0$, $G$ is a partial  isometry
and vice versa.

With $M$ a diagonal projection inserted, the equation of motion for $M$ simplifies to:

\begin{equation} \label{M proj}
M^{\ast} G^T \bar{G}  +  G \bar{M} \bar{G}  + G G^{\ast} M^{\ast} = 0
\end{equation}

If $M=I$ it simplifies further to:

\begin{equation}          \label{simplified}
G^T \bar{G} + G \bar{G} + G G^{\ast} = 0
\end{equation}

and whereas if $M$ is a diagonal projection of dimensionality $n$ and rank $m$ then the  bottom
$n-m$ rows of \ref{M proj} disappear.

We show below that  \ref{simplified} gives $G=0$ for $n=2$ while explaining the procedure in words
to make clear that this method can be applied to the general case of arbitrary $n$ and $m$:-
Consider the top left elements of $GG^{\ast}$ and $G^T \bar{G}$ where $n=2$:- all terms are positive
and all elements of the top row of $G$ are present. The equations containing the top left and
bottom right elements of $G\bar{G}$ may be combined as shown below to find that $G=0$.

For $n=2$ let

\begin{displaymath}
G=
\left(  \begin{array}{cc}
z & y \\
x & w
\end{array}   \right)
\end{displaymath}

the top left and bottom right equations are:

\begin{eqnarray*}
3 \vert z \vert^2 + 2 \vert x \vert^2 + y \bar{x} =0\\
3 \vert w \vert^2 + 2 \vert y \vert^2 + x \bar{y} =0
\end{eqnarray*}

and their combination is:
 
\begin{displaymath}
 3 \vert z \vert^2 + 3 \vert w \vert^2 +  \vert x \vert^2 +  \vert y \vert^2 + \vert x + y \vert^2
 \end{displaymath}
 
 which means that $x=y=w=z=0$ in other words $G=0$.

As mentioned, if $M=0$ then $G$ is a partial isometry. Equivalently, $GG^{\ast}$ and $G^{\ast}G$
are Murray-von-Neumann equivalent projections. And in the same way, we see for general $n$, $m$ the
bottom $n-m$ rows of equation \ref{simplified} show that the non-zero part of $G$ is a partial
isometry.

Notice the simple relationship between $M$ and $G$; when $M^{\ast}M$ and $G^{\ast}G$ are diagonal
projections and when $D_{F, vac}^2=I$, they are the orthogonal complement of one another. The
simultaneous matrix equations below simplify the statements: If $G=0$ then $M$ is a partial isometry
and vice versa, and if $M=I$ then $G=0$ and vice versa, and if $M$ is a diagonal projection and
if $D_{F, vac}$ is invertible, then $G$ is a partial isometry orthogonal to $M$.

\begin{eqnarray*} \label{G and M}
M ( M^{\ast} M + G^{\ast} G - I ) = 0 \\
G ( G^{\ast} G + M^{\ast} M - I ) = 0
\end{eqnarray*}

\subsection{Poincar\'e duality}

Given the standard model algebra and using Connes' realisation of Poincar\'e duality in
Noncommutative geometry,  we may write down the sign of the standard model Dirac operator. This is
just the matrix \ref{ad hoc} with eigenvalue `1' for each particle mass. An aim of the
previous work was to answer the question, `Is the standard model fermion mass matrix
(internal space Dirac operator) a solution to the additional Einstein's equations?' whereas here we
are considering the similar question, `Is the sign of the standard model internal space Dirac
operator a solution to the equations?' If the direction of this arrow is reversed, that is if the
$K$-theory group is identified for a given vacuum solution, then topological information
has been retrieved about the manifold from the minimisation procedure, and hence there would be a
mathematical reason for the choice of the algebra up to Morita equivalence. We begin by exploring
the relationship between vacuum solution and algebra.  We only study the first two summands of the
algebra, that is, the algebra for the Electroweak force. The matrix $M$ commutes with $\rho^c$ as
the Higgs has no colour charge and we do not involve colour charge at all in the remainder of this
article.

We study the standard model and one other solution.

Above a set of solutions was found where for a given projection $G$, the vacuum solution for $M$, or
$M_{vac}$ is determined via the simple relationship found in the equations of motion.  Let us
consider one such solution, namely the one in which $G=0$, that is, the one pertaining to the
standard model. Then the $M_{vac}^{\ast}$ is:

\begin{equation}        \label{M_v sm}
M_{vac}^{\ast} =
\left(   \begin{array}{ccc}
1             &          0          &   0  \\
0             &          1          &   0  \\
0             &          0          &   1  \\
0             &          0          &   0
\end{array}  \right)
\end{equation}

which is a rectangular matrix because the two direct summands of the chirality $\mathbb{Z}/2$-graded
Hilbert space have different dimensions. The basis may be labelled $(u_R,d_R,e_R)^T$. We can add a
final column of zeroes to $M_{vac}^{\ast}$ and the basis becomes $(u_R,d_R,e_R,\nu_R)^T$.

Recall that $[F] \in \rho(\mathcal{A}^{op})$ where $\rho(\mathcal{A})$ is given by \ref{rho}. The
opposite algebra is represented by  $J \rho (a) J^{-1}$. The generators of the $K$-homology
group are the homotopy classes of the minimal rank projections of the opposite algebra
$\mathcal{A}^o$, that is, $J \rho_L (p_1) J^{-1}$, which is the 2 by 2 unit matrix
$\mathrm{diag}(1,1)$ and $J \rho_R (p_2) J^{-1}$, which is given by the number $1$. The former is a
projection of the algebra of the quaternions $\mathbb{H}$ and the latter is simply a projection of
the complex numbers $\mathbb{C}$. So the vacuum solution \ref{M_v sm} is consistent with the element
$b^o$ of the opposite algebra being:

\begin{equation}
b^o =
\left(   \begin{array}{cccc}
\rho^{c~T}      &   0                    &   0                 &       0\\
0              &   \rho^{c~T}            &   0                 &       0\\
0              &   0                     &   q^T               &       0\\
0              &   0                     &   0                 &  \Lambda^T
\end{array}    \right)
\end{equation}

where $q$ is a quaternion and $\Lambda = \mathrm{diag}( \bar{\lambda},\lambda )$,
$\lambda \in \mathbb{C}$. Using the reality operator we find that an element $b$ of $\mathcal{A}$
can be given by:

\begin{equation}
b =
\left(   \begin{array}{cccc}
q              &   0                    &   0                 &       0         \\
0              &   \Lambda              &   0                 &       0         \\
0              &   0                    &   \rho^c            &       0         \\
0              &   0                    &   0                 &       \rho^c
\end{array}    \right)
\end{equation}

from which we see that the first two summands of the algebra may be: $\mathbb{H} \oplus
\mathbb{C}$.

Note that this is not a unique answer because the projection $\mathrm{diag}(1,1)$ is also in
$K_0(\mathbb{C})$.
 
A Fredholm module to be associated to a spectral triple must have algebra and Fredholm
operator compatibility such that the first order condition is satisfied. In order to check that a
spectral triple can be assigned to the Fredholm module, we check that the first order condition
is satisfied. It is satisfied because $G=0$ and $[M,\rho^c]=0$ where $M$ is given by the direct sum.

We can generalise this procedure by choosing any other projective solution for $G$. For example let
alternate eigenvalues of $G$ be non-zero beginning with the first eigenvalue zero. Then by the
orthogonal complement relationship we have the solution for $M_{vac}^{\ast}$:

\begin{equation}        \label{M_v c plus C}
M_{vac}^{\ast} =
\left(   \begin{array}{ccc}
1             &          0          &   0  \\
0             &          0          &   0  \\
0             &          0          &   1  \\
0             &          0          &   0
\end{array}  \right)
\end{equation}

and by the same procedure as above, we find that this solution is compatible with the algebra
$M_2(\mathbb{C}) \oplus \mathbb{C}$. $M_2(\mathbb{C})$ is Morita equivalent to $\mathbb{C}$ and so
the first two summands of the algebra of this spectral triple are Morita equivalent to $\mathbb{C}
\oplus \mathbb{C}$.

A point of clarification is that the two solutions considered above are unitary
(and homotopy\footnote{unitary and homotopy equivalence are the same thing for stable algebras})
equivalent, and so one expects them lead to the same topological invariants. However,
in the part to follow, we will restrict this homotopy freedom to separate `nodes'; one for each of
$\rho_L$, $\rho_R$ and $\rho^c$.

\subsection{Morse theory and the Witten complex}  \label{morse witten}

The  projection $\mathrm{diag}(1,1)$ is in the $K$-theory of both algebras $\mathbb{C}$ and
$\mathbb{H}$, but to obtain topological information from the vacuum solutions the relationship
between the vacuum solutions and the algebra must be one to one. We need
a reason why the solutions should correspond only to the generators of the
$K_0(\mathcal{A})$ group and not to any other element of the group. For example we need the matrix
$\mathrm{diag}(1,1)$ to be associated only with $\mathbb{H}$ and algebras Morita equivalent to it.

In this section we generalise a theorem involving Morse theory and the Witten complex to suggest a
method for finding topological information from the equations of motion about the $S^o$-real ($G=0$)
spectral triple pertaining to the standard model.  The theorem is that a chain complex called the
Witten complex, which is constructed from the critical points of the Morse function has the same
homology groups as the manifold that the Morse function is defined on. Atiyah and Bott have proven
that the Yang-Mills action is an equivariant Morse function and since the Higgs is a gauge field
component in the internal space direction, we ask whether the finite spectral action is also a
Morse function in a proposed noncommutative sense described below. To do this, we make a
straightforward generalisation of the theorem to noncommutative geometry (which comes down to
little more than the usual replacement of the commutative with the noncommutative algebra) by
proposing a noncommutative generalisation to the definition of Morse function, equivariant Morse
function (or Morse-Bott function), Hessian matrix and Witten complex.

A Morse function is a real-valued function on a (smooth) manifold $N$, $f:N \rightarrow \mathbb{R}$ such
that every one of its critical points is non-degenerate. The way to check for non-degeneracy is to
calculate the Hessian matrix of the critical point and if this has no zero eigenvalues then the
point is non-degenerate. Of course the Higgs vacuum is an entire three-sphere of non-degenerate
solutions, so in order to use Morse theory in physics problems where there is a gauge symmetry, the
equivariant Morse function was defined where the gauge symmetry is just divided out. (There
are more complicated cases).

The Hessian matrix of a Morse function $f$ for a critical point is given by:

\begin{equation} \label{Hessian of f}
   a_{ij} =   \bigg(  \frac{\partial^2 f}{\partial x_i \partial x_j}  \bigg)  \vert_c
\end{equation}

where $c$ is the critical point, the $x_i$ are the coordinates on $N$  and $i$ runs from 1 to the
dimension of $N$. The `index' of the critical point is the number of negative eigenvalues of this
matrix.

The action we want to consider is the finite spectral action with $G=0$:

\begin{equation}
   S = \mathrm{Tr} \big( (MM^{\ast})^2 + - 2MM^{\ast} \big)
\end{equation}

To proceed from here, we need to make some generalisations of the above definitions to the context
of noncommutative geometry. To begin with, the notion of a  critical \emph{point} is no longer
valid, and differentiating with respect to each of the commuting coordinates $x_i$ on $N$ goes over
to differentiating with respect to the matrix $M$. Consider the Fredholm operator $F$ ingredient of
the Fredholm module over a noncommutative $C^{\ast}$-algebra. Recalling that $F$ is the
generalisation to noncommutative geometry of an elliptic operator on a compact manifold in
commutative geometry, we understand that $F$ is parametrised by the underlying space pertaining to
the Fredholm module. Moreover, the homotopy classes of $F$, $[F] \in \rho(\mathcal{A}^{op})$. With
in mind the Fredholm module picture of the underlying space on which the Morse function acts, we
replace the $x_i$ with the $M$ as 2 by 2 matrices over $\mathbb{C}$. We use 2 by 2 matrices because
each direct summand of the standard model algebra is viewed as a node (not quite a point) upon which
the vector bundle is built, and the representation $\rho$ is separated into 2 by 2 `chunks'; one for
each algebra summand. Obviously the critical points will be replaced by the vacuum solutions
$M_{vac}$. Finally we need to write down a condition corresponding to \ref{Hessian of f} that will
give meaning to a noncommutative version of a Morse function, (or at least equivariant Morse
function) and check that the action $S$ satisfies that condition. We continue this chapter's focus
on the Fredholm module as the underlying space.

The index $i$ doesn't run very far because we are working in 0 dimensions and the action only
depends on $M$ and its hermitian conjugate. The Hessian matrix can only be as follows:

\begin{equation} \label{Hessian of S}
  a =   \bigg(  \frac{\partial^2 S}{\partial M^{\ast} \partial M}  \bigg)  \vert_c
\end{equation}

$S$ is real-valued, and we don't need to worry about smoothness as this is already covered in the
noncommutative geometry axioms. Differentiating the action $S$ with respect to $M$ and afterwards
with respect to $M^{\ast}$ and evaluating at $M=0$ produces  the Hessian of $S$. The resulting
matrix is: $MM^{\ast} + M^{\ast}M - I$ evaluated at the vacuum  solution which is that $MM^{\ast}$
and $M^{\ast}M$ are `initial' and `final' projections. Since in some solutions this matrix can have
zero eigenvalues, we differentiate twice with respect to $MM^{\ast}$, so that $MM^{\ast}$ becomes
the field to vary instead of $M$. The resulting matrix is $2I$, which has no zero eigenvalues. We
suggest then that $S$ is a noncommutative version of a Morse function. Since the degeneracies due
to the vacuum manifold exist and may be associated with the finding of 0 eigenvalues above, it may
be more accurate to designate the function as an equivariant Morse function in keeping with the
analogy with Yang-Mills theory.

In (commutative) Morse theory, the Witten complex is defined as follows (\cite{morse}). To begin,
the free abelian group $C_i$ generated  by the set of critical points of index $i$ is constructed.
If   $C_{i-1}$ is defined in the same way for index $i-1$, it is possible to define a map from $C_i$
to  $C_{i-1}$, that is, the boundary   map.    This defines a chain complex called the Witten
complex. It is a proven theorem that \emph{the homology groups of this chain complex are isomorphic
to the homology groups of the manifold.}

Classifying the vacuum solutions according to homotopy class, where each solution is
homotopic to one of the following two diagonal projections:

\begin{equation}        \label{}
M_{v1} =
\left(   \begin{array}{cc}
1             &          0   \\
0             &          1
\end{array}  \right) \qquad
M_{v2} =
\left(   \begin{array}{cc}
1             &          0    \\
0             &          0
\end{array}    \right)
\end{equation}

this is equivalent to dividing out the degenerate solutions as in equivariant theory, and to
considering the homotopy classes of the projections $(M M{\ast})_v$ where the action is varied with
respect to $MM^{\ast}$ instead of $M$.

Since the two solutions share the same Hessian, they have the same index and therefore the Witten
complex for this particular function has only one component $C_i$ so the homology groups are simply
$C_i$. Since the Witten chain complex in the case studied above is made out of only one term $C_i$,
the kernel of the boundary map is $C_i$, and the image of the boundary map for the next term in the
sequence takes the identity to the identity, therefore the homology group is simply given by $C_i$.
The abelian group $C_i$ is generated by the two projections $M_{v1}$ and $M_{v2}$, which is exactly
the $K$-homology group $K^0(\mathcal{A}^o)$, that is, the $K$-homology group of the underlying
Fredholm module. (The other elements of the abelian group arise from the other generations - of
which this framework implies there are an arbitrary number, and of course it is a mystery that we
have to stop at 3 - in analogy to the abelian group $\mathbb{Z}$ where $S^1 =
\frac{\mathbb{R}}{\mathbb{Z}}$.) The $K$-theory group of the standard  model algebra is isomorphic
to $\mathbb{Z} \oplus \mathbb{Z} \oplus \mathbb{Z}$ (though here we have only studied the first two
summands). Since one counts only a limited number of generations, the algebra is associated to the
punctured torus.

The two `critical nodes' can be identified with the first two direct summands (up to Morita
equivalence) on the internal space, one for $\mathbb{H}$ and the other $\mathbb{C}$, an example of
the noncommutative generalisation of the two point space, (left and right) as follows. We associate
$M_{v1}$ with left and $M_{v2}$ with right. So $M_{v1} \in \rho_L(\mathcal{A}^o)$ and $M_{v2} \in
\rho_R(\mathcal{A}^o)$. Removing $\nu_R$ from the Hilbert space and a row and column from $D_F$
means that the vacuum solution is: $M_Q=\mathrm{diag}(1,1)$ and $M_L=1$. These are the minimal rank
projections of $\mathbb{H}$ and $\mathbb{C}$ and Morita equivalent algebras. And by
Poincar\'e duality (as shown in the previous part), the noncommutative space can be described by the
quaternions over the right `node' and the complex numbers over the left.

\section*{Conclusions}

Although the solutions to the field equations calculated in our previous work yielded an unphysical
result, we have argued that the reasoning for their derivation deserves further attention and
we explained why \cite{smv} has highlighted an open question in the Noncommutative framework. A
solution was found to the equations of motion calculated previously with the `leptoquarks' held
constant whose features are compatible with the physical fermion mass matrix. Ironically, this
result provides a mathematical reason for the Yukawa couplings to
need fine-tuning.

The solutions to the field equations for $D_F$ both with and without the $S^o$-reality condition
were shown to be partial isometries. These were interpreted as the phase or sign of $D_F$. The
relationship between the vacuum solutions and the topology of the internal space was explored using
$K$-theory and $K$-homology and a method inspired by Morse theory to extract topological information
from the vacuum about the underlying Noncommutative space was developed. Instead of finding a
mathematical reason in the action principle for the \emph{geometry} of the standard model to be
what it is known to be by experiment, we have claimed that the vacuum provides information on its
\emph{topology} in terms of $K$-homology. Rather than the vacuum picking out just one of a myriad
of possible answers, all partial isometries are solutions. The study was limited to the first two
of three summands, that is, the Electroweak part of the standard model algebra.

We also note that with $D_F^2=I$, we can consider solutions in the case without
$S^o$-reality in which $M_{v2}$ is associated with its orthogonal complement matrix
$G_{v2}=\mathrm{diag}(0,1)$. Applying these solutions to the standard model basis, where $\nu_R=0$,
$G=0$ and the bottom row vanishes from $M_{v2}$ and the outcome is the same as above. Also, a
similar analysis can be carried out with $M$ having arbitrary dimensionalities and arbitrary
numbers of algebra direct summands. This involves either an unwanted prediction of new massless
particles, or cutting the matrices down as above to fit the standard model Hilbert space. For
example, in the dimensionality 3 case, there is an additional solution, $M_{v3}=\mathrm{diag}(1,1,1)
\in SU(3)$. Algebras with a greater number of summands cannot be identified using this method if
there are repeated summands, for example the algebra $\mathbb{H} \oplus \mathbb{H} \oplus
\mathbb{C} \oplus \mathbb{C}$ cannot be distinguished from $\mathbb{H} \oplus \mathbb{C}$ because
they have the same critical `nodes'.

\subsubsection*{Acknowledgements}

Advices from J. W. Barrett  and J. Zacharias were instrumental in the development of this research.
The Engineering and Physical Sciences Research Council (EPSRC) are recognised for their financial
support.

\section*{Appendix}

The equations of motion for one generation and three coloured quarks with `leptoquarks' held
constant. These were calculated for \cite{smv} using the computer package Maple.

\begin{equation}        \label{ac}
\bar{a}( 3 \vert a \vert^2 + 3 \vert c \vert^2 + 3 \vert b \vert^2 + \vert g \vert^2 +   \vert u
\vert^2 + \vert x \vert^2  + \vert h \vert^2 +\vert v \vert^2  + \vert y \vert^2  -3) +
3 \bar{c}\bar{b}d =0 \end{equation}

\begin{equation}        \label{bc}
\bar{b}( 3 \vert b \vert^2 + 3 \vert d \vert^2 + \vert g \vert^2 + \vert u \vert^2 +\vert x \vert^2
+\vert h \vert^2 +\vert v \vert^2 +\vert y \vert^2 + 3 \vert a \vert^2  -3) + 3 \bar{a}\bar{d}c =0
\end{equation}

\begin{equation}        \label{cc}
\bar{c}( \vert a \vert^2 + \vert c \vert^2 + \vert d \vert^2  -1) + \bar{a}\bar{d}b =0
\end{equation}

\begin{equation}        \label{dc}
\bar{d}( \vert b \vert^2 + \vert d \vert^2 + \vert c \vert^2  -1) + \bar{c}\bar{b}a =0
\end{equation}

\begin{equation}        \label{qc}
\bar{q}( \vert q \vert^2 + \vert s \vert^2 + \vert r \vert^2 + \vert g \vert^2 +\vert u \vert^2 +\vert x \vert^2 + \vert j
\vert^2 -1) + \bar{r}(\bar{h}g +\bar{v}u   +\bar{y}x+ \bar{l}j + \bar{s}t) =0
\end{equation}

\begin{equation}        \label{rc}
\bar{r}( \vert r \vert^2 + \vert t \vert^2 + \vert q \vert^2 + \vert h \vert^2 +\vert v \vert^2 +\vert y \vert^2   + \vert l \vert^2 -1) + \bar{q}(\bar{g}h +\bar{u}v  +\bar{x}y+ \bar{j}l + \bar{t}s) =0
\end{equation}

\begin{equation}        \label{sc}
\bar{s}( 3\vert j \vert^2 + \vert q \vert^2 + \vert s \vert^2 +\vert t \vert^2 + \vert g
\vert^2 +\vert u \vert^2  +\vert x \vert^2 + \vert l \vert^2 -1) + \bar{t}(\bar{h}g +\bar{v}u  +\bar{y}x+ 2\bar{l}j+
\bar{q}r) =  0  \end{equation}

\begin{equation}        \label{tc}
\bar{t}( 3\vert l \vert^2 + \vert r \vert^2 + \vert t \vert^2 + \vert s \vert^2 + \vert h
\vert^2 +\vert v \vert^2  + \vert y \vert^2  + \vert j \vert^2 -1) + \bar{s}(\bar{g}h +  \bar{u}v  +\bar{x}y+ 2\bar{j}l +
\bar{r}q)  =   0  \end{equation}

\begin{equation}        \label{jc}
\bar{j}( 3\vert s \vert^2 + \vert j \vert^2 + \vert g \vert^2 +\vert u \vert^2 +\vert x \vert^2 +  \vert l \vert^2 + \vert t
\vert^2  + \vert q \vert^2 -1) + \bar{l}(\bar{r}q + 2\bar{t}s +
 \bar{g}h   + \bar{u}v + \bar{x}y )=0  \end{equation}

\begin{equation}        \label{lc}
\bar{l}( 3\vert t \vert^2 + \vert h \vert^2 + \vert v \vert^2 +\vert y \vert^2 + \vert j \vert^2 +\vert l \vert^2 + \vert r
\vert^2  + \vert s \vert^2 -1) + \bar{j}(\bar{q}r + 2\bar{s}t +
\bar{h}g  + \bar{v}u + \bar{y}x  )=0  \end{equation}

 \end{document}